\documentclass[twocolumn,showpacs,amsmath,amssymb,prl]{revtex4}

\usepackage{graphicx}
\usepackage{bm}

\begin{document}

\preprint{APS/123-QED}

\title{Commensurability effects in overlap Josephson junctions
       coupled  \\ with a magnetic dots array}%

\author{S.~N.~Vdovichev}
\author{S.~A.~Gusev}
\author{Yu.~N.~Nozdrin}
\author{A.~V.~Samokhvalov}
\author{A.~A.~Fraerman}
\affiliation{%
Institute for Physics of Microstructures, Russian Academy of
Sciences, 603950, Nizhny Novgorod, GSP-105, Russia}
\author{E.~Il'ichev}
\author{R.~Stolz}
\author{L.Fritzsch}
\affiliation{%
Institute for Physical High Technology, P.O. Box 100239, 07702
Jena, Germany}

\date{\today}                   
\begin{abstract}
Experimental observation of the strong influence of an array of
ferromagnetic nanodots on the 
critical current of a short overlap Josephson junction is
reported. Pronounced commensurability effects are detected due to
the presence of the additional peaks in the magnetic field induced
diffraction pattern. The changes in the Fraunhofer pattern of the
Josephson junctions are account for by the formation of Abrikosov
vortices trapped in the electrodes which induce a phase
inhomogeneity in the junction area.
\end{abstract}

\maketitle


The role played by inhomogeneities in
long Josephson junctions is known.
In typical experimental realizations, periodic defects 
of the barrier thickness provide for a spatial modulation of the
critical current density $j_c$ which results in sharp peaks of the
critical current $I_c$ on variation of the external magnetic field
$H$ \cite{JJCommens}.
These ordered peak structures are observed
at fields $H$ for which the number of enclosed magnetic flux quanta in the
junction is an integer multiple of the number of defects, which is
known as commensurability.

Over the last years, Josephson junctions with phase inhomogeneities
have been intensively investigated.
In earlier studies the properties of Josephson junctions with
Abrikosov vortices (AVs) pinned in the vicinity of
the barrier were discussed by several
authors \cite{AV-JJ,Golubov:87}.
Recently investigations of unconventional Josephson junctions
with a spatially alternating sign of the critical
current have attracted renewed attention.
The presence of Josephson phase discontinuities
results in an unusual current--phase relation
\cite{CurrPhase},
a highly anomalous non-Fraunhofer $I_c(H)$ dependence
\cite{nonFraunh},
and a spontaneous generation of fractional Josephson vortices
\cite{SemiFlux}.
Although regular commensurability
peaks similar to the found in
\cite{JJCommens}
are quite expected
for unconventional junctions, observing them is difficult
due to the effect of randomness in the spatial distribution of
$j_c$.

On the other hand, superconducting (SC) thin films coupled
with arrays of magnetic nanodots show clear commensurability
effects in the measurements of magnetization,
critical current, and resistivity
\cite{MDots}.
Enhanced pinning effects have already been demonstrated
when the AV 
lattice and the magnetic dots
array were matched.
However, in real SC films the effects
are hidden by a strong intrinsic pinning
and are observable only in a narrow temperature
range near $T_C$.%
\begin{figure}
\centerline{\includegraphics[width=0.8\linewidth]{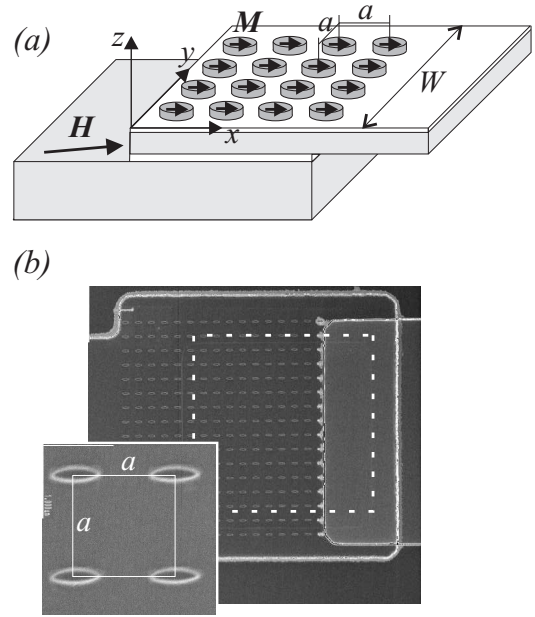}} %
\caption{\label{fig:1} (a) A regular square array of elongated
magnetic particles on top of the SC electrode of the overlap
Josephson junction. (b) SEM image (top view) of the setup. The
dashed line indicates the junction region. The thick top electrode
is shown by grey. The inset shows the elementary cell of the dots
array.}
\end{figure}

Therefore, it is quite natural to combine these activities, namely
to attach an array of magnetic dots to a Josephson junction.
Indeed, recently such hybrid ferromagnet/superconductor (FS)
systems have been successfully prepared and studied.
The system consists of a Josephson junction,
formed by an edge contact of two SC films
and coupled to a chain of
magnetic nanodots
\cite{Edge-MD}.
The stray magnetic fields of the dots partially penetrate
into the junction region and induce a spatially modulated
Josephson phase difference $\varphi$.
As a result, we have observed strong dependence of the
diffraction pattern $I_c(H)$ on the preliminary
magnetizaion of the particles.

%
\begin{figure}
\centerline{\includegraphics[width=0.9\linewidth]{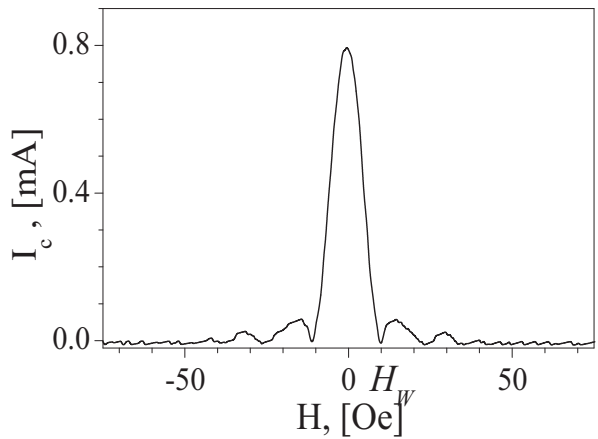}} %
\caption{\label{Fig:2} Dependence $I_c(H)$ for the junction
without the magnetic dots.}
\end{figure}
In this Letter, we propose and implement
a hybrid FS system
which allows to study commensurability effects in
short unconventional Josephson junctions.
In this context the term "unconventional" means:
these junctions exhibit a phase variation along
a contact on a scale which is much smaller
than the Josephson penetration depth.
Due to a strictly periodic Josephson phase modulation,
we succeeded in observing pronounced commensurability
peaks in $I_c(H)$ dependence.

The investigated short overlap Josephson junction
coupled with a
magnetic dots array is shown in Fig.1a.
The dots are separated
from the top SC electrode by an insulating layer,
which ensures no proximity effect
and that the interaction between
the junction and the dots has
only magnetic character.
Since the area occupied by the dots and the barrier region
are far from each other and are separated by thick SC
layer, we can neglect the penetration of the stray field
of the dots through the edges of the junction.
On the other hand, magnetic fields that reach the barrier
by means of Meissner screening supercurrents
does not create a Josephson phase difference.
So, modulation of the Josephson phase
like considered in Ref.
\cite{Edge-MD}
can be omitted.

The basic idea of the experiment is to form in the top SC
electrode a lattice of AVs 
by cooling the junction
through the transition temperature $T_C$ in the
magnetic field of an array of magnetic dipoles consisting of
single--domain magnetic dots with in-plane magnetization.
Single flux quanta with opposite polarity are induced in the SC
film at the north and south poles of the each magnetic dot
and can be considered as a vortex-antivortex pair
\cite{MD-VAV}.
The trapped vortex--antivortex pairs generate a
built-in magnetic
field in the barrier,
which induces a Josephson phase difference.


For our experiments, a series of ${\rm Nb/Al-AlO_x/Nb}$ overlap
junctions were fabricated, employing conventional technology
\cite{Stolz:99} with the area $W^2\! \sim\, 20\times 20\, {\rm \mu
m^2}$. The essential feature of the junctions is the thin top
electrode (${\rm Nb}$) with
a thickness is about 
$30\, {\rm nm}$, covered with a $50\, {\rm nm}$ ${\rm SiO}$
dielectric.
The junction has the critical current
density of about $j_c\approx 100\, {\rm A/cm^2}$
which corresponds to the Josephson length
$\lambda_J \approx 25\, {\rm \mu m}$
\cite{Likharev-Dyn}.
On top, the square array (lattice period $a = 1.4\, {\rm \mu m}$)
of elongated Co nanodots was fabricated by magnetron sputtering
and electron beam lithography \cite{Fraerman:02}. Figure~1b shows
a secondary electron microscopy (SEM) image of the system under
study. The Co dots have lateral dimensions of $650\, {\rm nm\,
(easy\, axis)}\times 280\, {\rm nm}$ and a thickness of $55\, {\rm
nm}$. The magnetic state of the dots was monitored by magnetic
force microscopy (MFM) at room temperature. After magnetization
along the easy axis all dots are in an uniformly magnetized
remanent ground state. The inset of Fig.~3 shows the MFM image of
the array of preliminarily magnetized nanodots. The dipole stray
fields characteristics are clearly visible. A minor variation in
the contrast for different particles is explained by the small
tilt angle between sample and scanning planes. Demagnetization
converts these magnetic particles into a 
multivortex state.
Measurements of the critical current $I_c$ as a function of
the external magnetic field $H$, applied in the plane of the
junction, were made using computer--controlled current sweeps at
${\rm 4.2\, K}$. The criterion for $I_c$ is the detection of a
voltage,
typically about $5\, {\rm \mu V}$. %

Figure~2 shows the ordinary Fraunhofer dependence $I_c(H)$ of the
Josephson junction
before the magnetic particles were sputtered:
there is the critical current maximum at zero field and
oscillations of $I_c$ which diminish in amplitude with increasing
field $H$.
The measured period of the diffraction pattern %
$H_W = \Phi_0 / \Lambda W \simeq 11.3\, {\rm Oe}$ enables the
determination of the effective depth of the magnetic field
penetration into the SC electrodes
$\Lambda \approx 0.1\, {\rm \mu m}$, %
where $\Phi_0$ is the magnetic flux quantum.
The same Fraunhofer dependence was observed
when the magnetic dots are mainly in the vortex state.

Qualitatively a different diffraction pattern is observed if the
array of the uniformly magnetized nanodots is attached to the
junction. Figure~3 shows the field--dependent critical current
$I_c(H)$ of the junction coupled with the ordered lattice of
magnetic dipoles. One can see that the typical value of the
Josephson current through junction is noticeably suppressed in
comparison with the case before. The symmetry of the diffraction
pattern with respect to the external magnetic field inversion is
broken: $I_c(-H)\neq I_c(H)$. There is also a set of observable
critical current maxima for high magnetic fields $|H| > H_W$. The
two  dominant side peaks of the $I_c(H)$ curve are clearly
observed at values of the field $H_{+}=100\,{\rm Oe}$ and
$H_{-}=-60\,{\rm Oe}$. The relative position of $H_\pm$ peaks
correspond to the addition/removal of one flux quantum $\Phi_0$
per period $a$, and the following commensurability condition is
satisfied to a good accuracy:
\begin{equation}\label{eq:1}
    \Phi_a(H_{+}) - \Phi_a(H_{-}) \approx \Phi_0,
\end{equation}
where $\Phi_a(H) = H S_a$ is the value of the magnetic
flux of the field $H$ through an elementary cell
$S_a=\Lambda\times a$.
\begin{figure}
\centerline{\includegraphics[width=0.9\linewidth]{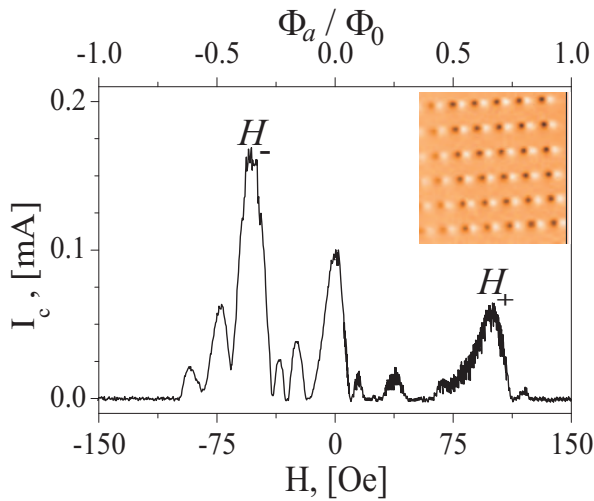}} %
\caption{\label{Fig:3} Dependence $I_c(H)$ for junction coupled
with the array of the uniformly magnetized nanodots. 
The insert shows the MFM images of the part ${10\,\times\,10\,
{\rm \mu m^2}}$ of the ordered lattice of dipoles.}
\end{figure}
%


The studied system can be represented by a square Josephson junction
in which the gauge--invariant phase difference
$\varphi(\mathbf{r})$ depends on a two-dimensional vector
$\mathbf{r}=(x,y)$ lying in the junction plane.
The junction occupies the area $0 \le x,y \le W$.
The size $W$ is small in
comparison with the Josephson penetration depth ($W < \lambda_J$),
and self--field effects of the Josephson current are negligible.
Magnetostatic calculations show that for a typical
value of the saturation magnetization
$M_s = 800\, {\rm Oe}$ and for the parameters taken
from the experiments, the stray field of both
poles of the in-plane magnetized dot creates a
(positive or negative) flux $\Phi_s > \Phi_0$
through the surface of the top SC electrode.
So, in accordance with the criterion proposed in
\cite{MD-VAV},
we assume that each magnetic dot creates
a pair of opposite vortices which pierce
the top electrode of the junction.
The phase difference $\varphi^M(\mathbf{r})$, created by the array
of magnetic dots is determined only by the positions of
vortices trapped in the top electrode of the junction, and obeys the
following equations:
\begin{eqnarray}
    &\triangle\varphi^M&\; =\, 0\; , \label{eq:2} \\
    {\rm curl}_z(\!\!&\nabla\varphi^M&\!\!) = 2\pi\sum_{n,m} %
       \left[ \delta(\mathbf{r}-\mathbf{r}_{nm}^+)%
             -\delta(\mathbf{r}-\mathbf{r}_{nm}^-) %
       \right]                               \label{eq:3}
\end{eqnarray}
with the following boundary conditions at the edges of the
junction:
\begin{equation} \label{eq:4}
    \partial_x \varphi^M \bigg\vert_{x=0,W} = 0, \quad %
    \partial_y \varphi^M \bigg\vert_{y=0,W} = 0.
\end{equation}
Here $\mathbf{r}_{nm}^{\pm}=(a n \mp d,a m)$ are the coordinates
of vortices ($+$) and antivortices ($-$) created by the square
($a\times a$) array of the  uniformly magnetized particles
$\mathbf{M} = M \mathbf{x}_0$, having a size equal to $2d$ in the
easy direction
\cite{MD-VAV}.

The critical current across the junction
is given by
\cite{Josephson:65ap}
\begin{equation}\label{eq:5}
    I_c = j_c \left\vert{\int\limits_0^W dx dy\, %
       \sin\left[\varphi(x,y)\right]} \right\vert,
\end{equation}
where the phase difference
\begin{equation}\label{eq:6}
    \varphi(x,y)=\varphi^M(x,y)
        +\frac{2\pi\Lambda}{\Phi_0} \left(H_x y - H_y x\right)
\end{equation}
depends both on the phase difference $\varphi^M$ and on the
external magnetic field %
$\mathbf{H} = H_x \mathbf{x}_0 + H_y\mathbf{y}_0$ %
applied in the junction plane. By using the described model, we
performed simulations of the dependence of the critical current
$I_c$ on the external magnetic field $\mathbf{H}$ for various
values of the relation $d/a$ and the different values of the angle
$\alpha$ between the $\mathbf{H}$ and $\mathbf{M}$ direction.
Figure~4 shows the results of a simulation of the diffraction
pattern $I_c(\phi_a)$ ($\phi_a=\Phi_a(H_x) / \Phi_0$) for the
parameters $W = 14 a$ and $d = a / 3$, which are close to those of
the samples in use.
%
\begin{figure}
\centerline{\includegraphics[width=0.9\linewidth]{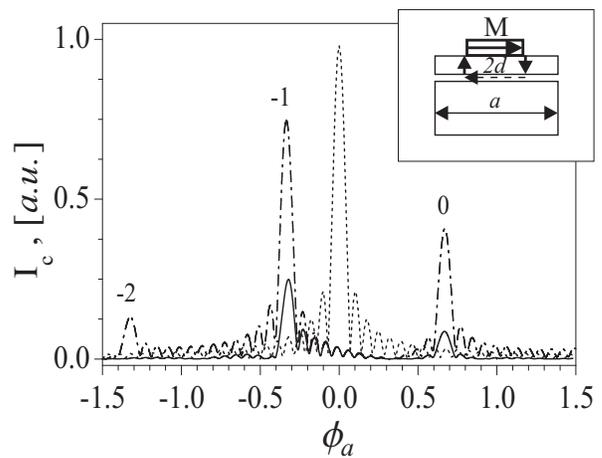}} %
\caption{\label{Fig:4} Dependence of $I_c(H)$ for a junction when
the vortex--antivortex pairs form the ordered square lattice ($W =
14 a$, $d = a/3$). The dash-dotted curve corresponds to
$\alpha=0^o$, the solid curve to $\alpha=9^o$. The dotted line
represents the Fraunhofer pattern in the absence of
vortex--antivortex pairs in the top electrode. The numbers near
the peaks denote the corresponding values of $n$ in (\ref{eq:7}).
The inset schematically shows the elementary cell of the junction
with the attached magnetic dot above. The vortex--antivortex pair is
represented by the vertical arrows near the opposite poles of the
nanoscale magnet.}
\end{figure}
It is seen that $I_c(\phi_a)$ oscillates with a slightly varying
amplitude when the flux $\phi_a$
increases.
The period of the
oscillations $\triangle\phi_a$ is determined by the junction width
$W$: $\triangle\phi_a = a / W$. Strong additional peaks occur at
\begin{equation}\label{eq:7}
    \phi_a[n] = n + \phi_d,\quad n=0,\pm1,\pm2\ldots ,
\end{equation}
where $\phi_d = 2 d /a$. One observes that the positions of the
two dominant peaks calculated from the 
condition
(\ref{eq:7}) for $n=-1,\,0$ and $d = a /3$ ($\phi_a[-1] \simeq
-0.33$, $\phi_a[0] \simeq 0.66$) is in good agreement with the
experimental data (see Fig.~3 to compare). The fixed shift of the
peaks $\phi_d$ can be explained by the presence of
the built-in field
$\mathbf{H}_{in} = H_{in}\,\mathbf{x}_0 = - H_a (2d/a)\,\mathbf{x}_0$,
which creates
the average
gradient of the phase difference:
$\left\langle \partial_y \varphi^M\right\rangle %
    = - q_a (2d / a)$.
Here $H_a = \Phi_0 / S_a$ is the field of
one flux quantum enclosed in the cell area, and $q_a = 2\pi /a$ is
the inverse lattice constant.
The peak for $n=0$ occurs when the
$x$-projection of the external magnetic field $\mathbf{H}$
overcomes the built-in field $\mathbf{H}_{in}$:
$H_x + H_{in} = 0$.

The position of the dominant peaks is determined
by commensurability between the periodic
modulation of the Josephson current due to magnetic dots
and the imposed modulation owing to the external
magnetic field $\mathbf{H}$.
The maximum contribution from the oscillating term in $j = j_c
\sin\left[\varphi(x,y)\right]$ to the total current across the
junction $I$ corresponds to
the field $\mathbf{H}$ for
which the the spatial Josephson current wave
\begin{equation}\label{eq:9} 
    j \sim \sin(\mathbf{q}\, \mathbf{r}),\quad
    \mathbf{q} = (q_x,\, q_y)
    =\frac{2\pi \Lambda}{\Phi_0} %
    \left(\mathbf{H} + \mathbf{H}_{in}\right)
\end{equation} 
has the same space harmonics as the phase factor $\varphi^M(x,y)$.
If the angle $\alpha$ is small so as $q_y \ll q_a$,
the resonant condition between the
wave number $q_x$ and the inverse lattice constant $q_a$
\begin{equation}\label{eq:10}
    q_x = n\, q_a,\quad n=0,\pm1,\pm2\ldots
\end{equation}
defines the values of the external magnetic field $H_x$ for which
the commensurability peaks of $I_c(H)$ arise:
\begin{equation}\label{eq:11} 
    H_x = n H_a - H_{in}
\end{equation} 
By taking into account that $\phi_a = H_x / H_a$, it is easy to
rewrite the last equation in the form of condition (\ref{eq:7}).
Note that the considerable suppression of the characteristic
critical current observed in the experiments can be explained by
misalignment of the external magnetic field $\mathbf{H}$ with
respect to the easy axis of the magnetic dots.

So, most of the qualitative peculiarities of the experimentally
observed diffraction pattern can be consistently explained in the
framework of the model, when the array of the uniformly magnetized
nanodots creates an ordered lattice of vortex--antivortex pairs in
the top electrode of the junction. However, the critical current
dependence $I_c(H)$ in the vicinity of $H=0$ (see Fig.~3) is not
consistent with the result of  the simulations presented at
Fig.~4. This disagreement between theory and experiment can be
explained by the junction design. There is a certain area near the
edge of the junction with a thick top electrode (about $200\, {\rm
nm}$), when the magnetic dots and, thus, a periodic modulation of
the phase difference are absent. This part of the junction has an
ordinary Fraunhofer pattern like shown in Fig.~2, and is
responsible for the central peak of $I_c(H)$ in Fig.~3.

In summary, we measured commensurability peaks in $I_c(H)$ of the
short overlap Josephson junctions
coupled to an array of the
magnetic nanodots.
The position and the amplitude of the observed
peaks conform to the proposed model in which the
vortex--antivortex pair are created by cooling the junction
through the transition temperature $T_c$ in the stray magnetic
fields of the in-plane magnetized dots. Thus, we have made the
indirect confirmation that single flux quanta with opposite
polarity are induced in the SC layer at the opposite poles of the
in-plane magnetic dot \cite{MD-VAV}. The experimental results show
that the vortex lattice is quite perfect despite of the intrinsic
pinning inherent in SC films. We observed a strong variation of
the maximum critical current which depends on the magnetic state
of the particles. This gives us the possibility to control the
Josephson phase in the hybrid FS systems.

\begin{acknowledgments}
We are indebted to
A.S. Mel'nikov %
and %
I.A. Shereshevskii
for useful discussions, and to G.L. Pakhomov and B.A. Gribkov for
assistance in the preparation of the samples and MFM measurements.
The work was supported by the RFBR (\#03-02-16774) and the program
RAS "Quantum Macrophysics".
\end{acknowledgments}


\end{document}